\documentclass[12pt]{article}
\usepackage{latexsym}
\usepackage{amssymb}
\usepackage{amsmath}
\usepackage{graphicx}

\newcommand{\footnoteremember}[2]{
  \footnote{#2}
  \newcounter{#1}
  \setcounter{#1}{\value{footnote}}
}
\newcommand{\footnoterecall}[1]{
  \footnotemark[\value{#1}]
}

\begin{document}
\title{{\small \bf{GRAVASTAR SOLUTIONS WITH CONTINUOUS PRESSURES AND EQUATION OF STATE}}}
\author {{\small A.~DeBenedictis \footnote{Pacific Institute for the Mathematical Sciences,
Simon Fraser University site, Burnaby, British Columbia, V5A 1S6, Canada \hspace{0.4cm}
(adebened@sfu.ca)} \footnoteremember{sfuphys}{Department of Physics, Simon Fraser University,
Burnaby, British Columbia, V5A 1S6, Canada}}
\hspace{0.3cm}{\small D.~Horvat \footnoteremember{zagrebaff}{Department of Physics, Faculty of Electrical
Engineering and Computing, University of Zagreb, Unska 3, HR 10000 Zagreb, Croatia }\footnoterecall{sfuphys}}
\hspace{0.3cm}{\small S.~Iliji\' c \footnoterecall{zagrebaff}}
\hspace{0.3cm}{\small S.~Kloster \footnote{Center for Experimental and Constructive Mathematics,
Simon Fraser University, Burnaby, British Columbia, V5A 1S6, Canada}}
\hspace{0.3cm}{\small K.~S.~Viswanathan \footnoterecall{sfuphys}}}

\date{{\small June 15, 2007}}
\maketitle

\begin{abstract}
\noindent We study the gravitational vacuum star (gravastar)
configuration as proposed by \cite{[CFVis]} in a model where the
interior de Sitter  spacetime segment is continuously extended to
the exterior Schwarzschild spacetime. The multilayered structure
of \cite{[MM-1]} - \cite{[MM-3]} is replaced by a continuous
stress-energy tensor at the price of introducing anisotropy in the
(fluid) model of the gravastar. Either with an ansatz for the
equation of state connecting the radial $p_r$ and tangential $p_t$
pressure or with a calculated equation of state with
non-homogeneous energy/fluid density, solutions are obtained which
in all aspects satisfy the conditions expected for an anisotropic
gravastar \cite{[CFVis]}. Certain energy conditions have been
shown to be obeyed and a polytropic equation of state has been
derived. Stability of the solution with respect to possible axial
perturbation is shown to hold.
\end{abstract}

\vspace{3mm}

\section{{\normalsize INTRODUCTION}}
Recently it has been proposed by several authors that
objects other than black holes could be formed
by gravitational collapse of a massive star. Black hole
horizons introduce number of theoretical problems and a
consensus of solutions of those problems has not yet been reached \cite{[MM-1]}, \cite{ref:chls}.

One proposal, which was initiated recently by Mazur and Mottola
(M-M) \cite{[MM-1]}-\cite{[MM-3]} is the so called ``gravastar".
In this scenario, quantum vacuum fluctuations are expected to
play a non-trivial role in the collapse dynamics. A phase
transition is believed to occur yielding a repulsive de Sitter
core which helps balance the collapsing object thus preventing
horizon (and singularity) formation \cite{ref:gliner},
\cite{ref:chls}. It is expected that this transition occurs very
close to the limit $2m(r)/r=1$ so that, to an outside observer,
it would be very difficult\begin{scriptsize}\end{scriptsize} to
distinguish the gravastar from a true black hole. Since this
proposal, different versions of the original gravastar model have
appeared with all variety of ingredients. For example, it has
been shown how a gravastar structure can form from a Born-Infeld
scalar field such as that predicted by low energy string theory
\cite{ref:BF} or supported from non-linear electrodynamics
\cite{ref:loboem}. In the multi-layer structure of the M-M model
with de Sitter core and asymptotical Schwarzschild outside region,
additional features have been added. In \cite{[Carter]} gravastar
solutions have been studied in a generalized
(Reissner-Nordstr\"om) exterior and solutions of the model
stemming from the original Mazur-Mottola model have been
analyzed. Gravastar type solutions in the context of solutions to
Einstein's equations with tube-like cores have also been recently
considred \cite{ref:zasla}. Pioneering in-depth studies of
spherically symmetric systems with deSitter asymptotics may be
found in \cite{ref:dymnik1} - \cite{[Dymnikova]}. More recently
several papers have appeared discussing limits on gravastars and
how to distinguish them from black holes \cite{ref:BN},
\cite{ref:cr}.

It has recently been shown \cite{[CFVis]} that the gravastar
configuration (see the next section for the elaborated physical
model of it) has to have anisotropic pressures which in addition
should obey some of the energy conditions of General Relativity.
In this context one would ideally construct such a spherically
symmetric model of an anisotropic fluid with a corresponding
equation of state. The model, by the definition of the gravastar,
should not possess a horizon, it should to be stable and its
(anisotropic) pressures and density ideally would not violate
energy conditions. Certainly this last requirement is somewhat
relaxed in the sense that configurations constructed in this way
will violate some of the (usual) energy conditions from its very
initial definition. Full reviews of gravastar models may be found
in  \cite{ref:catthesis} and \cite{ref:fabthesis}.

\begin{figure}[h!t]
\begin{center}
\includegraphics[bb=2247 2474 2494 2702, scale=0.8, clip, keepaspectratio=true]{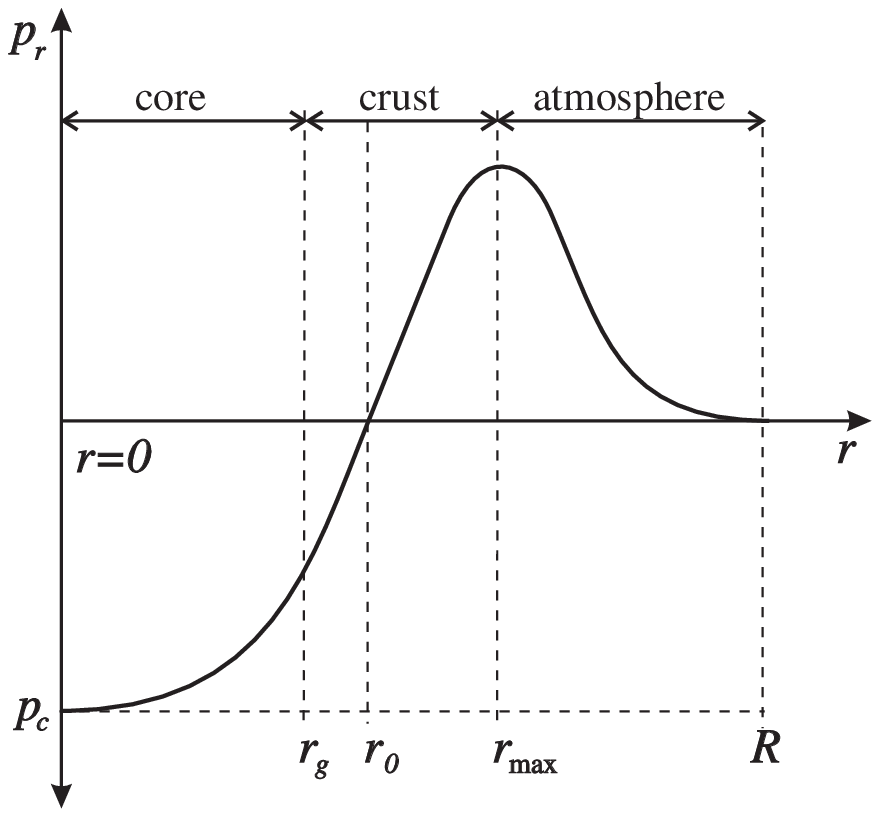}
\caption{{\small Sketch of the gravastar as proposed by Cattoen, Faber and Visser
\cite{[CFVis]}.}} \label{fig:1}
\end{center}
\end{figure}

In this paper we present  solutions for the gravastar as proposed by
Cattoen, Faber and Visser (CFV) \cite{[CFVis]}. In their paper an attractive sketch (see
Fig.\ref{fig:1} drawn here for the convenience of the reader) is given
as a guide to all the future gravastar model builders. In the next section a
general description of gravastars as anisotropic fluid spheres is
presented.

In section 3\ we describe the evolution of the anisotropic gravastar
model and one of the solutions (model 1) which reproduces the CFV sketch of the
anisotropic gravastar. The value of the surface redshift is calculated while a more elaborate analysis of this important quantity is relegated to a future work. Section 4\  is devoted to the solution (model 2)
which
starts from the construction of an
equation of state and consistently solving
the remaining generalized Tolman-Oppenheimer-Volkoff equation. Stability of this
solution with respect to axial perturbations is calculated.
Finally, we conclude with comments and possible extensions of the anisotropic
gravastar solutions.

\section{{\normalsize FROM ISOTROPY TO ANISOTROPY}}

Einstein's field equations, being exceedingly complicated because of
their nonlinear character, have one most important closed form solution
outside a spherical star of total mass $M$, namely the Schwarzschild metric. Its
line element in curvature coordinates is

\begin{equation}
ds^2=-\left(1-\frac{2M}{r}\right)dt^{2} +\frac{dr^{2}}{\left(1-{2M}/{r}\right)}+ r^{2}\,d\Omega^{2}\;, \label{eq:schwmet}
\end{equation}
where $d\Omega^2:=d\theta^{2}+\sin^2\theta\,d\phi^2$.

For the interior of the star one has to choose a ``physically
reasonable" stress-energy tensor. One attractive possibility is
to use the perfect fluid model of matter where
\begin{equation}
T^{\mu}_{\ \nu}=\left(\rho+ p\right)u^{\mu}u_{\nu}+p \, \delta^{\mu}_{\;\nu}\;, \label{eq:pertmunu}
\end{equation}
with $\rho$ and $p$ the energy density and pressure respectively
in the co-moving frame of the fluid, and $u^{\mu}$ being the fluid
four-velocity. In the static case, which is studied here, the
interior metric may be written as
\begin{equation}
ds^2=-e^{\nu(r)}\,dt^2+e^{\lambda(r)}\,dr^2+r^2\,d\Omega^{2}. \label{eq:gensphmet}
\end{equation}

The Einstein equations, $ G^{\mu}_{\ \nu}=8\pi T^{\mu}_{\ \nu}\;$,
give a system of equations:
\begin{subequations}
\begin{align}
e^{-\lambda}\left(\frac{\nu'}{r}+\frac{1}{r^2}\right)-\frac{1}{r^2}&=8\pi
p\;, \label{eq:E1} \\
e^{-\lambda}\left(\frac12 \nu''-\frac14 \lambda'\nu'+
\frac14 (\nu')^2+\frac{\nu'-\lambda'}{2r}\right)&=8\pi p\;, \label{eq:E2} \\
e^{-\lambda}\left(\frac{\lambda'}{r}-\frac{1}{r^2}\right)+\frac{1}{r^2}&=8\pi
\rho \;. \label{eq:E3}
\end{align}
\end{subequations}
These are supplemented with the conservation law $T^{\mu}_{\;
\nu; \mu}=0$, which in this case yields only one non-trivial
equation:
\begin{equation}
T^{\,\prime\,1}_{\;\;\;\;1}+\frac{2}{r}\left[1+\frac{r}{4}
\nu^{\prime}\right] T^{1}_{\;1} -\frac{1}{2} \lambda^{\prime}
T^{0}_{\;0} -\frac{2}{r}T^{2}_{\;2}=0\;. \label{eq:conslaw}
\end{equation}

Elimination of the function $\nu(r)$ from the above
under-determined system leads to a convenient form of the
conservation equation i.e. the  Tolman-Oppenheimer-Volkov (TOV)
\cite{[OppV]}, \cite{[Tol]} equation:
\begin{equation}
\frac{dp(r)}{dr}=-\frac{[\rho(r) +p(r)][m(r)+4\pi p(r)r^3]}{r^2[1-2m(r)/r]}\;, \label{eq:TOV}
\end{equation}
with
\begin{equation}
m(r):=-4\pi \int_0^r\,T^{0}_{\;0}(\tilde{r}) \:\tilde{r}^{2} \,d\tilde r. \label{eq:massdef}
\end{equation}
One may also specify an equation of state which
relates pressure and density.
If, for simplicity, we adopt at this stage a constant density profile function (with
built-in boundary conditions)
\begin{equation}
\rho(r)=\left\{
\begin{array}{l}
\rho_0\qquad \text{for $r<R$}\\
0 \qquad \;\:\text{for $R<r$}
\end{array}
\right.
\end{equation}

then it turns out to be an oversimplification which leads to
analytic integration of (\ref{eq:TOV}) and the corresponding field equations but pressure and density do
not obey the energy conditions
which are analogous to the requirement of mass  positivity in the
Newtonian mechanics.

With the isotropic fluid and the above homogeneous energy density
static solutions are allowed for objects with a restricted total
mass $M$ to radius $R$ ratio i.e. $2M/R\leq 8/9$
\cite{ref:Buchdahl}.

\subsection{{\small ANISOTROPY}}
\medskip
The idea of anisotropy in the spherically symmetric geometry was
perhaps first introduced by G. Lema\^{i}tre \cite{[Lemet]} and
suggested by Einstein (as quoted  in Ref. \cite{[Lemet]}). The
limiting case of $p_r\to 0$ is mentioned there and the remaining
transversal pressure was said to be enough to support a (stable)
sphere (see also \cite{[Florides]}). Further development has
brought different refinements of the original anisotropy notion
(see the papers \cite{ref:aniso0} - \cite{[MahCha]} for studies
and further references).

The perfect fluid requires that the pressure in the interior of a
star be isotropic, leading to calculations of isotropic polytropes for
descriptions of objects like white dwarfs or neutron stars.
Another option giving more freedom to the equation of state within
the spherical symmetry is the introduction of the
stress-energy tensor which is anisotropic in its
principal pressures. The anisotropy is sometimes (spontaneously)
produced by extending the notion of a (perfect) fluid to
phenomenological models including eg. scalar fields, or strongly
interacting matter, although it is not known how large this
anisotropy may be in realistic models.

The stress-energy tensor for an anisotropic matter/fluid distribution
is given by
\begin{equation}
T^{\mu}_{\ \nu}=\left(\rho+p_{t}\right)u^{\mu}u_{\nu}
+p_{t}\,\delta^{\mu}_{\;\nu}
+\left(p_{r}-p_{t}\right)s^{\mu}s_{\nu}\;,\label{eq:anisotmunu}
\end{equation}
where $p_r$ is the radial pressure in the co-moving frame and
(again) due to the spherical symmetry, the angular components are
identified and are denoted as transversal pressures, $p_t$. The
vector $s^{\mu}$ is orthogonal to the fluid four-velocity
($s^{\mu}u_{\mu}=0$).

In the gravastar model, the
core interior is assumed to be given by a de Sitter solution so the
appropriate pressure/density ratio value equal to minus one should be implemented
as an initial condition for the density profile function, as well as
a corrector to an equation of state connecting the pressure and density.

For the moment we assume constant energy density $\rho_{0}$. In addition
to the (energy) density in a prescribed form,
a relation between the radial $p_r$ and the
tangential $p_t$ pressures should be given. An ansatz for the
anisotropy measure,
\begin{equation}
\Delta:=\frac{p_t-p_r}{\rho}\; , \label{eq:anisopar}
\end{equation}
will be used following
the hints given in \cite{[CFVis]}. Bounds on the anisotropy measure
are calculated and are expressed in terms of the ``compactness" $2m(r)/r$,
so our ansatz has the following form:
\begin{equation}
{\Delta}=\frac{\alpha^2}{12}\frac{2m(r)}{r}\;. \label{eq:mod1delta}
\end{equation}
The constant $\alpha^2/12$ will simplify the (numerical) calculation and
it is a measure of anisotropy for this version of the gravastar model.
The TOV equation now assumes the following form:
\begin{equation}
r^2p'_{r}(r)=\frac{4\pi r^3}{9}\left[\alpha^2\rho_0^2-
\frac{3[\rho_0+p_r(r)][\rho_0+3p_r(r)]}
     {1-8r^2\pi\rho_0 /3}\right].
\end{equation}
with the initial condition $p_r(0)=p_0$, requiring also that
$p_r(0)=p_t(0)$.

The radial pressure $p_r(r)$ is given by
\begin{equation}
p_r(r)=\frac{\rho_0}{3}\left[
 -1+\alpha\sqrt{1-\mu(r)}\left(-1+
 \frac{2(\alpha-2)}
      {\alpha-2+(\alpha+2)\exp[\alpha(-1+\sqrt{1-\mu(r)})]}
      \right)\right].
\end{equation}
where the ``compactness" is  $\mu (r)=2m(r)/r=8\pi r^3\rho_0/3$, while the transversal pressure can be obtained through
(\ref{eq:mod1delta}), and is given by
\begin{align}
p_t(r)=&\frac{\rho_0}{12}\left\{4\left[
 -1+\alpha\sqrt{1-\mu(r)} \right. \right. \nonumber \\
&\left.\left. \cdot\left(-1+
 \frac{2(\alpha-2)}
      {\alpha-2+(\alpha+2)\exp[\alpha(-1+\sqrt{1-\mu(r)})]}
      \right)\right]+\mu\alpha^2\right\}.
\end{align}
For all $\alpha>0$, both pressures start at $-\rho_0$ in
the centre of the gravastar. At $\mu=1$ (if one would allow
such a density profile) the values are $p_{r}=-\rho_{0}/3$, $p_{t}=\frac{\rho_{0}}{3}\left(\alpha^{2}-1\right)$.
Only for values $\alpha>4.111\,72$, does the radial pressure reach
positive values before returning back into the negative pressure
region. It is also worth noting that this simple model does not lead to
an equation of state, since the energy density does not change
within the star.

As previously mentioned, this oversimplified model with the
constant (energy) density function as given above cannot
reproduce the sketched Figure \ref{fig:1}. i.e. it cannot provide
a gravastar in the proposed model without the surface layer
structure as required by junction conditions. The most obvious
extension is to use perhaps the same ansatz for the equation of
state connecting the behaviour of radial and tangential pressures
as given in (\ref{eq:mod1delta}). For the density one takes some
non-homogeneous distribution which will give a more complex
dependence of $p_r(r)$ and lead to a possible solution (see the
model 1 below).

Certainly the ansatz could be replaced by a calculated equation of state
 and introduce again an inhomogeneous density
distribution (see the model 2 below). On these two possibilities the new
results of this paper are based.

\subsection{{\small ENERGY CONDITIONS}}
In order to have a reasonably realistic radial pressure component we demand the (most natural)
weak energy condition (WEC) be satisfied everywhere:
\begin{equation}
\rho(r)\geq 0\quad\text{along with}\quad \rho(r)+p_{r}(r) \geq 0\quad\text{and}\quad \rho(r)+p_{t}(r) \geq 0\;. \label{eq:WEC}
\end{equation}
 Also, we require
\begin{equation}
\rho(r=0)> 0
\end{equation}
and, since stability of the fluid/gravastar requires that $\rho(r)$ must not
increase outwards,
\begin{equation}
\frac{d\rho(r)}{dr}\leq 0\:.
\end{equation}

The above WEC obviously implies:
\begin{equation}
\rho(r)+p_r(r)\geq 0\quad\text{and}\quad \rho(r)+p_{t}(r) \geq 0\;, \label{eq:NEC}
\end{equation}
the  so called null
energy condition (NEC).
There is also a dominant energy condition (DEC) which requires that
\begin{equation}
\rho(r)\geq 0\quad\text{and}\quad p_r(r)\in [-\rho(r),+\rho(r)]
\quad\text{and}\quad p_t(r)\in [-\rho(r),+\rho(r)], \label{eq:DEC}
\end{equation}
and which implies the other two energy conditions.

Another commonly studied energy condition is the strong energy condition (SEC) which states that, for our static system,
\begin{equation}
\rho+p_{r}\geq 0,\quad \rho+p_{t}\geq 0 \quad\text{and}\quad \rho+p_{r}+2p_{t} \geq 0. \label{eq:SEC}
\end{equation}
Since gravastars possess a deSitter core, it is not possible to satisfy this energy condition.

Within the normal range of the gravastar it could be reasonable to
propose that the speed of sound shall not exceed 1 (speed of
light). This requirement is reasonable to apply in the region
where one could expect that the unusual physics govern most of the
physical processes (i.e. where the matter possesses the least
exotic behaviour). From the sketch in Figure \ref{fig:1} this is
expected to be in the gravastar atmosphere, so
\begin{equation}
v_s^{\,2}=\frac{dp}{d\rho}\Big\vert_{\text{atm.}}\leq 1. \label{eq:soundlimit}
\end{equation}

In this region one should be able to derive a polytropic equation of state which should be
expressed by parameters which are in accordance to the requirement
involving the speed of sound (see next Section).

\section{{\normalsize MODEL 1: THE ANSATZ AND AN
INHOMOGENEOUS DENSITY DISTRIBUTION}}

In this section the solution of the gravastar non-layered model
(i.e. the model sketched in Figure \ref{fig:1} \cite{[CFVis]}) is
sought by prescribing an equation of state relating the pressures.
i.e. an equation which connects the radial $p_r(r)$ and
tangential $p_t(r)$ pressure. An improved ansatz of the form
\begin{equation}
\tilde{\Delta}=\Delta\frac{\rho}{\rho_{0}} \label{eq:newdelt}
\end{equation}
is assumed, where $\Delta$ is as given in (\ref{eq:mod1delta}). The (inhomogeneous) density function profiles will be
chosen to be either of  the exponential form or a polynomial form.

The gravastar is a static configuration with a prescribed
behaviour of the (radial and tangential) pressure. As sketched in
Figure \ref{fig:1} the radial (and tangential) pressure
originates at $r=0$ with the initial value conforming with the de
Sitter interior definition $\left(p_{r}
(r=0)=p_{t}(r=0)=-\rho(r=0) \right)$.
Investigations concerning a possible analytic form of the (energy)
density have produced stringent bounds on the allowed functional
behaviour.

In the first approach to the gravastar model we start from the anisotropic
TOV equation. Following the requirement imposed on the density profile we
will be using a simply behaving function of the form
\begin{equation}
\rho(r,r_0)=\rho_0e^{-(r/r_0)^n}\;, \label{eq:mod1rho}
\end{equation}
where $r_0$ has an appropriate dimension and $n$ is chosen to be a
positive integer. Motivations for the exponential form of $\rho$
may be found in \cite{[Dymnikova]},
\cite{[Mbonye]}, where they chose $n=3$, which simplifies some calculations. The
values of the central density, $\rho_0$, and of the fall-off constant, $r_{0}$, may not be set arbitrarily, since one must make sure
that the compactness does not exceed unity. Alternatively, one
may require a certain total mass of the configuration, and set the
upper limit to the compactness within the star. Then the values
for the parameters $\rho_0$ and $r_{0}$  would follow. In this (and
the following) model we will restrict our (numerical) procedures
to a general total mass $M$ and $r_{0}$ by
keeping the corresponding quantities as ratios.

When requirements of anisotropy are being imposed we use the
anisotropic fluid stress-energy tensor
(\ref{eq:anisotmunu}) $T^{\mu}_{\ \nu}=\text{diag}(-\rho,\,
p_r,\,p_t,\,p_t)$. As before, we have from the equations
(\ref{eq:E1})-(\ref{eq:conslaw})
\begin{equation}
m(r)=4\pi \int_0^r \rho(\tilde r)\,\tilde r^2\,d\tilde r
\end{equation}
and the generalized TOV equation
\begin{align}
\aligned
\frac{dp_r(r)}{dr}&=-[\rho(r)+p_r(r)]\frac{m(r)+4\pi p_r(r)r^3}{r^2[1-2m(r)/r]}
+2\frac{p_t(r)-p_r(r)}{r}\\
&=-[\rho+p_r(r)]\frac{m(r)+4\pi p_r(r)r^3}{r^2[1-2m(r)/r]}
+2\rho(r)\frac{\tilde{\Delta} (r)}{r}
\endaligned \label{eq:anisoTOV}
\end{align}
where we have introduced the anisotropy parameter $\tilde{\Delta}(r)$ from above.

\begin{figure}[ht]
\begin{center} \includegraphics[bb=0 0 374 180, scale=0.8, clip, keepaspectratio=true]{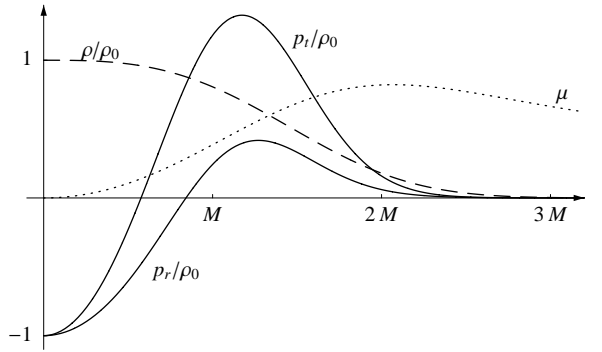}
\caption{{\small The gravastar model
with the energy density profile $\rho=\rho_0\,\exp[-(r/r_0)^n]$
and anisotropy $(p_{t}-p_{r})/\rho=\alpha^{2}\;(\rho/\rho_{0})\;\mu/ 12$:
radial (lower solid line) and transversal (upper solid line) pressures,
energy density (dashed line) and the compactness (dotted line).
$M$ is the total mass of the configuration.
In this example the parameters are $n=3$,
maximal compactness within the gravastar $\mu_{\mbox{max}}=0.822$,
and $\alpha=6.69$. $M=1$ here so, for a body of mass 10 km, distances are measured in units of 10 km.}} \label{fig:dympress}
\end{center}
\end{figure}

In Figure \ref{fig:dympress} the (numerically obtained) solutions
of the transversal $p_t$ and radial $p_r$ pressures are depicted
when the modified ansatz  given in (\ref{eq:newdelt}) is used as
an input for the equation of state. Also the density profile
$\rho(r)$ given in (\ref{eq:mod1rho}) (decreasing dashed line) as
well as $\rho_0\cdot \mu(r)=2\rho_0\cdot m(r)/r$ for $\rho_0=1$
(initially increasing dashed line) is presented.

The radial pressure $p_r(r)$
as well as energy density satisfy all the energy conditions
assumed to be valid for gravastar solutions.

Instead of insisting on numerical details we want here to show
that the proposed gravastar structure could be found also with
other energy density profile functions which also follow general
requirements for a gravastar. By choosing another independent
profile we will show that solutions found in this paper are not
just artefacts of a very specifically chosen energy density profile
but a more general feature of prescribed initial conditions
within the framework of general relativity.

\subsection{{\small MODEL 1b}}
As the next example of the gravastar model construction we will
use the density profile function of the form
\begin{equation}
\rho(r)=\rho_0[1-(r/R)^n];\quad n\geq 2 \;,\label{eq:polyrho}
\end{equation}
and with the ansatz given in (\ref{eq:mod1delta}).
Again the values of the central density, $\rho_0$, and of $R$ are adjusted to make sure that the
compactness does not exceed unity.
As an example we used $n=4$ in (\ref{eq:polyrho})
to construct a gravastar of total mass $M=1$
with maximal compactness within the gravstar $\mu_{\mathrm{max}}=0.93$.
For these parameters we obtained $\rho_0=0.037$ and $R=2.31$,
and the compactness at the surface $\mu(R)=0.864$. If $M$ is measured in decametres, the central density in this model is approximately $4\times 10^{24}$ kg/Dm$^{3}$, approximately 1000 times neutron star density. The gravitational repulsion of the near deSitter type core region is supporting such densities.

\begin{figure}[h!t]
\begin{center} \includegraphics[bb=0 0 370 180, scale=0.8, clip, keepaspectratio=true]{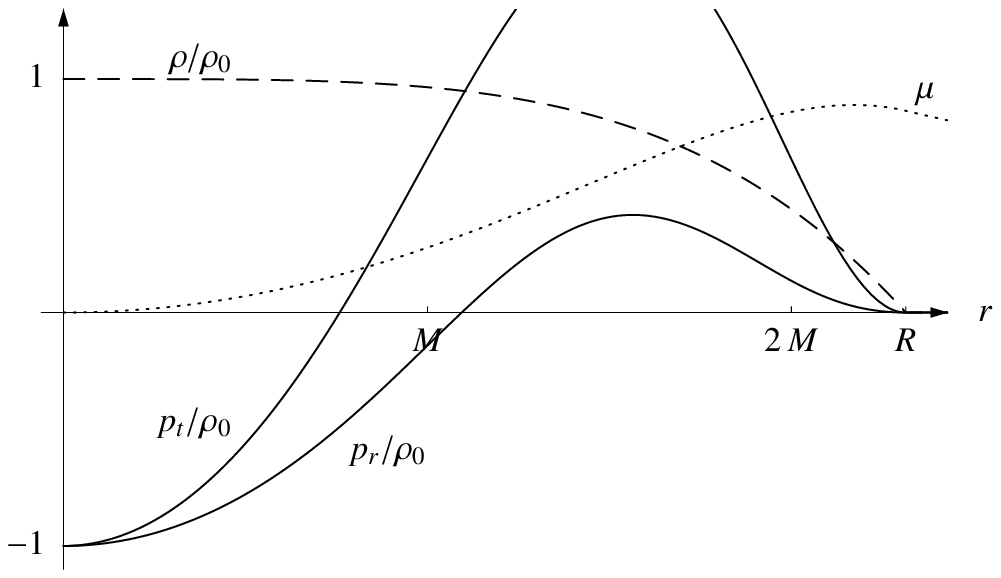}
\caption{{\small The gravastar model
with the energy density profile $\rho=\rho_0(1-(r/R)^n)$
and anisotropy $(p_{t}-p_{r})/\rho=\alpha^{2}\;(\rho/\rho_{0})\;\mu/ 12$:
radial (lower solid line) and transversal (upper solid line) pressures,
energy density (dashed line) and the compactness (dotted line).
$M$ is the total mass of the configuration.
In this example the parameters are $n=4$,
maximal compactness within the gravastar $\mu_{\mbox{max}}=0.889$,
and $\alpha=6.10$.  Surface compactness is $\mu(R)=0.864$.}}
\label{fig:polypress}
\end{center}
\end{figure}

By adjusting the value of $\alpha$ we
could obtain a solution in which both pressures vanish at $r=R$,
thus obtaining a gravastar.
The appropriate value was $\alpha=6.10$,
while for lower/upper values the pressures end up at
negative/positive values at the surface of the gravastar, such
solutions were not considered any further (although the tangential pressure and energy density need not vanish at the surface). The pressures, the
density profile and the compactness of this model are shown in
Fig.~\ref{fig:polypress}.

\subsection{{\small THE EQUATION OF STATE AND SURFACE REDSHIFT}}
\medskip

Apart from the energy condition relations among the components of
$T^{\mu}_{\ \nu}$, an equation of state for the matter described
by $T^{\mu}_{\;\nu}$ should exist. An equation of state could be
either given in the sense that it is prescribed as an intial
condition under which the Einstein field equations should be
solved or derived as {\it a posteriori} calculated law which
should be in accordance with the expected characteristics of a
physical system described by the $T^{\mu}_{\ \nu}$ content.

The dependence of the (radial) pressure on the energy density is usually,
in the perfect fluid regime, expressed in a mathematical form
\begin{equation}
p=\kappa \rho^{1+1/n_{p}}=\kappa \rho^{\gamma} \label{eq:polytrope}
\end{equation}
which represents the polytrope, and $n_{p}$ is a polytropic index
\cite{[Wein]}) .

\begin{figure}[h!t]
\begin{center}
\includegraphics[bb=0 0 370 180, scale=0.8, clip, keepaspectratio=true]{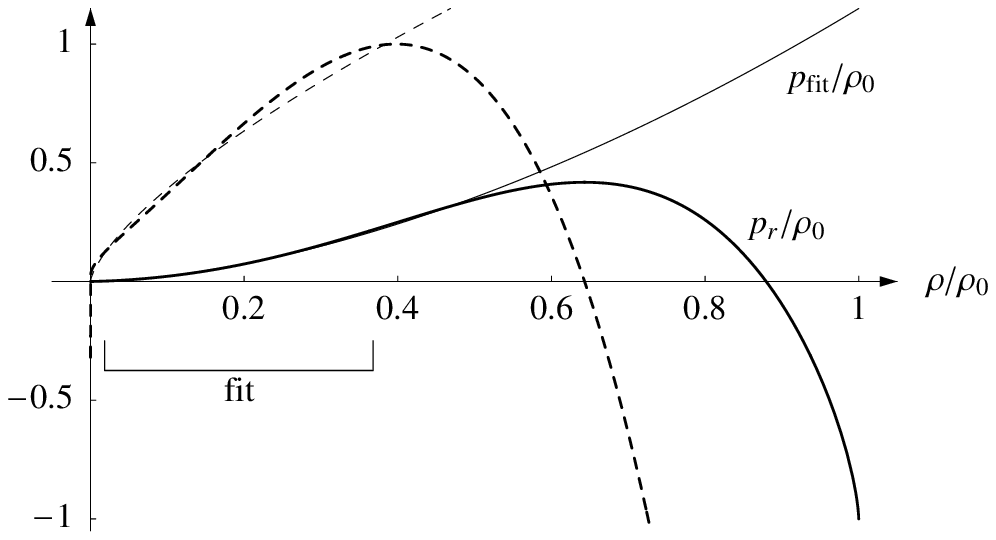}
\caption{{\small The equation of state
resulting from the solution shown in Fig.~\ref{fig:dympress}:
radial pressure (thick solid line) and its derivative
with respect to the energy density (thick dashed line).
The polytropic fit computed in the range indicated by the marker
and extrapolated over the whole range of densities:
radial pressure (thin solid line) and its derivative (thin dashed line).
The parameters of the polytropic fit are $\kappa=9.16$ and $\gamma=1.70$.
The atmosphere corresponds to the left region of the graph.}} \label{fig:dymstate1}
\end{center}
\end{figure}

According to the graph of the radial pressure vs. density (in the
mirrored presentation - see Figure \ref{fig:dymstate1}) it is clear that
the equation of polytrope could/should be calculated only in the
``atmosphere" of the gravastar (see Figure \ref{fig:1}), where
radial pressure and density follow the usual behavior of a more
standard fluid and the equation of state is assumed to relate
energy density to radial pressure, as is commonly employed when dealing with anisotropic fluids. In the polytropic regime the speed of sound respects the bound $dp/d\rho < 1$. The specific models constructed here are designed to illustrate the maximum compactness allowed by the speed of sound causality condition (\ref{eq:soundlimit}). The parameters can easily be adjusted to produce models where the speed of sound is much less than unity.

One method of determining a star's properties is the measurement of its surface redshift, $Z$, of spectral lines produced in the star's photosphere. It is defined by the fractional change ($\Delta\lambda$) between observed ($\lambda_{0}$) and emitted ($\lambda_{e}$) wavelength compared to emitted wavelength:
\begin{equation}
Z:=\frac{\Delta \lambda}{\lambda_{e}}=\frac{\lambda_{0}}{\lambda{e}}-1\: , \label{eq:redshift}
\end{equation}
and, according to our notation, its surface value becomes
\begin{equation}
Z_{R}=e^{-\nu(R)/2}-1\,. \label{eq:gravred}
\end{equation}
In the context of the presented model calculations it is important to recall that anisotropy affects the (surface) redshift so that for the static perfect fluid sphere the surface redshift is not larger then $Z_s=2$ \cite{ref:Buchdahl}, \cite{ref:aniso2} whereas in \cite{ref:aniso2}, \cite{ref:aniso3b}, and references therein, the {\it maximal\/} surface redshift for anisotropic sphere is found to be 3.842. The model 1a (see Fig. 2) provides the value $Z_s=1.230$ (with maximum compactness $\mu_{\text{max}}=0.822$). The model 1b (Fig. 3), with $\mu_{\text{max}}=0.889$ and surface compactness $\mu_{\text{s}}=\mu(R)=0.864$ gives the value $Z_s=1.712$. Both values are well within expected values for a realistic (neutron star) object.

\begin{figure}[h!t]
\begin{center}
\includegraphics[bb=0 0 370 181, scale=0.8, clip, keepaspectratio=true]{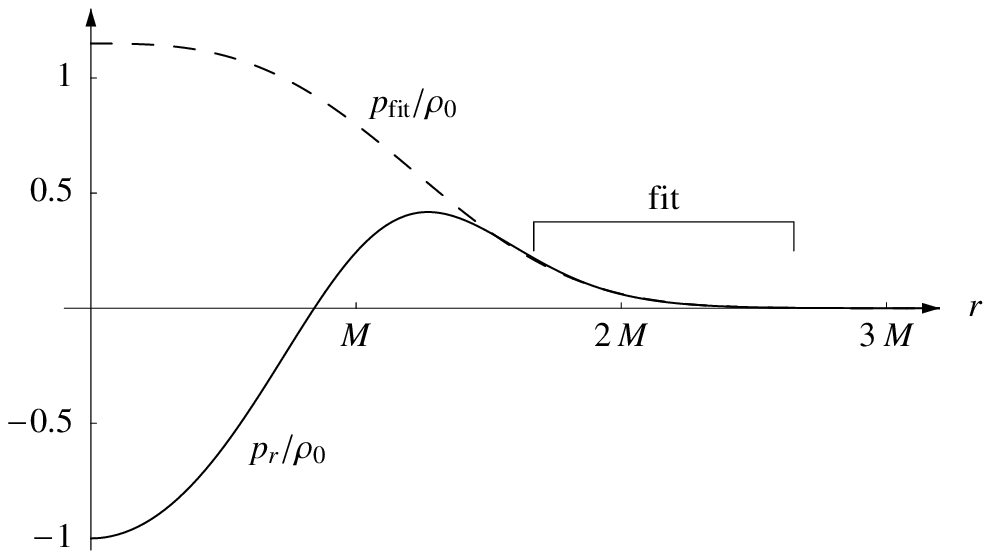}
\caption{{\small The radial pressure of Fig.~\ref{fig:dympress} (solid line)
and the polytropic fit radial pressure (dashed line) vs.\ $r$. }} \label{fig:dymstate2}
\end{center}
\end{figure}

\begin{figure}
\begin{center}
\includegraphics[bb=0 0 370 181, scale=0.8, clip, keepaspectratio=true]{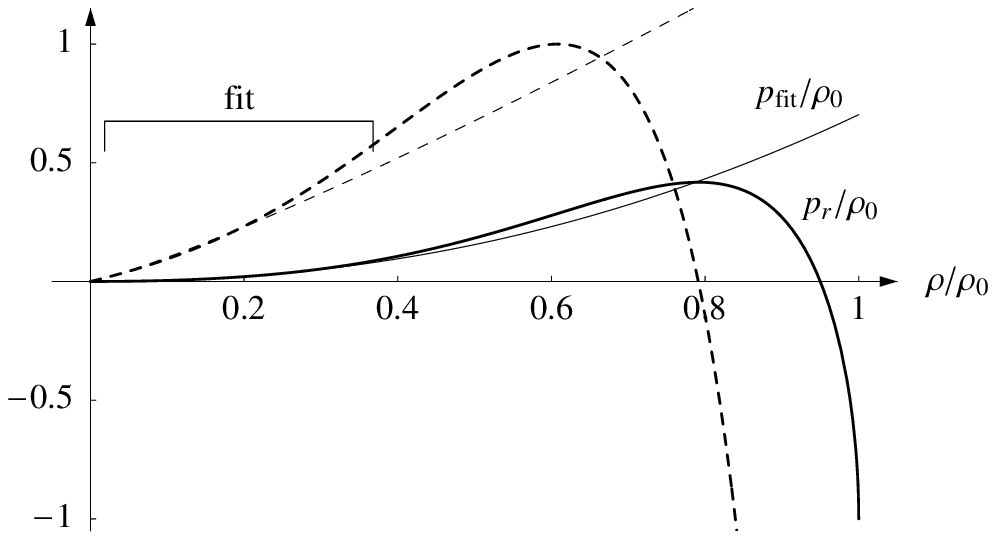}
\caption{{\small The equation of state
resulting from the solution shown in Fig.~\ref{fig:polypress}:
radial pressure (thick solid line) and its derivative
with respect to the energy density (thick dashed line).
The polytropic fit computed in the range indicated by the marker
and extrapolated over the whole range of densities:
radial pressure (thin solid line) and its derivative (thin dashed line).
The parameters of the polytropic fit are $\kappa=38.0$ and $\gamma=2.18$.
The atmosphere corresponds to the left region of the graph.}}
\label{fig:polystate1}
\end{center}
\end{figure}

\begin{figure}
\begin{center}
\includegraphics[bb=0 0 370 180, scale=0.8, clip, keepaspectratio=true]{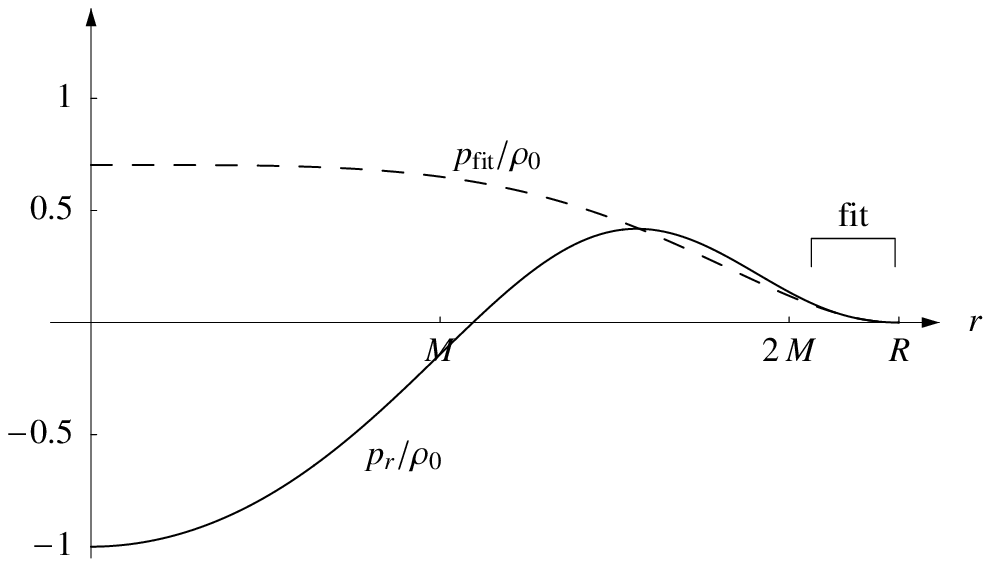}
\caption{{\small The radial pressure of Fig.~\ref{fig:polypress} (solid line)
and the polytropic fit radial pressure (dashed line) vs.\ $r$.}} \label{fig:polystate2}
\end{center}
\end{figure}
\clearpage

\section{{\normalsize MODEL 2}}
Here we select a density profile of the form (\ref{eq:mod1rho}). Specifically,
for numerical calculations, we chose
\begin{equation}
\rho_{0}=\frac{1}{16\pi M^{2}},\;\;\; r_{0}=(12)^{1/3}M\,, \;\;\; M=6000\, \mbox{m}\;. \nonumber
\end{equation}
The energy density is related to the radial pressure via a Mbonye-Kazanas (MK)
equation of state \cite{[Mbonye]}:
\begin{equation}
p_r(\rho)=\left[s-(s+1)\left(\frac{\rho(r)}{\rho_0}\right)^m\right]
\left(\frac{\rho(r)}{\rho_0}\right)^{1/n}\cdot \rho(r)\;.
\label{eq:MKEOS}
\end{equation}

The MK equation of state possesses several desirable features
(with appropriately chosen parameters), namely: the speed of
sound is less than one in the atmosphere and the WEC and DEC are satisfied.
The model considered here utilizes (\ref{eq:MKEOS}) with the parameters $m=2$
and $n=1$. The tangential pressure will then be given by the
anisotropic TOV equation (\ref{eq:anisoTOV}).

\begin{figure}[h!t]
\begin{center}
\includegraphics[bb=80 4 373 180, scale=0.8, clip, keepaspectratio=true]{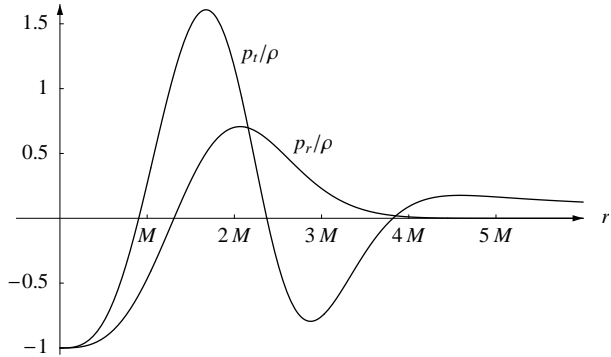}
\caption{{\small The DEC with radial pressure $p_r(r)/\rho(r)$
and with tangential pressure $p_t(r)/\rho(r)$.
The parameters are: $s=2.2135$ and total mass $M=6\,\mathrm{km}$.}} \label{fig:8}
\end{center}
\end{figure}

In Figure \ref{fig:8} we plot $p_{r}/\rho$ (solid) and $p_{t}/\rho$ (dashed) in order to study the DEC.

From Figure \ref{fig:8} we see that there exists DEC violation
with respect to the tangential pressure. This violation occurs in
the crust of the gravastar, where the physics is expected to be
``exotic'' and generally cannot be avoided as noted in
\cite{[CFVis]}. Note that in a model constructed via this method,
the DEC violation is minimal and the DEC can be respected in the
outer layer of the star.

\begin{figure}[h!t]
\begin{center}
\includegraphics[bb=80 4 373 180, scale=0.8, clip, keepaspectratio=true]{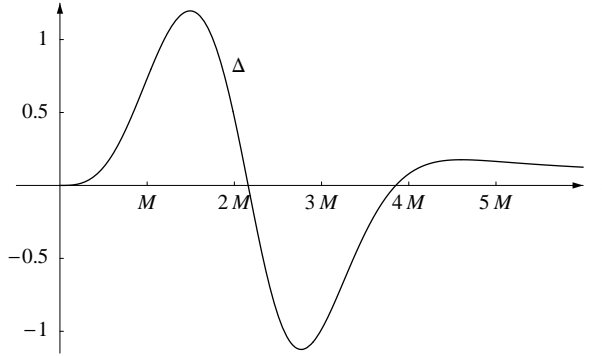}
\caption{{\small The anisotropy parameter, $\Delta=(p_t(r)-p_r(r))/\rho(r)$, for model 2.}} \label{fig:9}
\end{center}
\end{figure}

The expression for the anisotropy measure $\Delta$ (which is now calculated and not prescribed as in model 1), although in closed form, is not simple. Instead, for the specific model
considered here, we plot it in Figure \ref{fig:9}. Notice from this figure that $\Delta$ vanishes in
several locations. It should be possible at these points to patch the solution to a perfect
fluid yielding a regular star-like structure in the outer region.

Following the calculation in the previous model, the specific model constructed here possesses a surface redhift of $Z_{R}=0.4\:$.

\subsection{\small AXIAL STABILITY}
Here we analyze stability of this model against axial
perturbations. Stability of  gravastar models with thin-shells
against spherical perturbations has been analyzed in several
papers \cite{ref:viswilt}, \cite{ref:lobo}. The model of Lobo
\cite{ref:lobo} considers the interesting scenario where the
gravastar like object is formed from the gravitational
condensation of dark energy, believed to be responsible for the
current acceleration of the universe. The physical motivation for
considering dark energy stars may be found in \cite{ref:chap}.

We concentrate here on the issue of axial perturbations, which
allow for perturbative rotations. This is useful as, in a
realistic astrophysical collapse scenario, one would expect the
star to possess some amount of angular momentum. For details the
interested reader is referred to the book by Chandrasekhar
\cite{ref:chandra} and the paper by Dymnikova and Galaktionov
\cite{ref:dymg}. Before proceeding we shall establish some
notation.
\begin{equation}
g_{00}=-e^{\nu(r)}=-e^{-\lambda(r)}e^{\Gamma(r)}, \label{eq:g00}
\end{equation}
where
\begin{equation}
e^{-\lambda(r)}=1-\frac{2m(r)}{r} \nonumber
\end{equation}
and
\begin{equation}
{\Gamma(r)}= \int_{\tilde{r}=0}^{r}\,h(\tilde{r})\,d\tilde{r}\;,
\nonumber
\end{equation}
\begin{equation}
h(r)=\frac{8\pi r^{2}\left(\rho+p_{r}\right)}{r-2m(r)}  \label{eq:h}
\end{equation}
(these will be used below).

A more general line element may be written as
\begin{equation}
ds^{2}=-e^{\nu}\,dt^{2} + e^{\lambda} \, dr^{2} +
r^{2}\,d\theta^{2} +r^{2}\sin^{2}\theta\left[d\phi -\omega\,dt
-q_{2}\,dr -q_{3}\,d\theta\right]^{2} . \label{pertmet}
\end{equation}
By comparison with (\ref{eq:gensphmet}), the unperturbed metric
has $\omega=q_{2}=q_{3}=0$. Axial perturbations correspond to
these becoming non-zero. Our analysis is to linear order in these
quantities.

We study the form of the equations asymptotically, in the region
where $e^{\lambda}=-e^{-\nu}$, within a time rescaling. With some
restrictions, this will be sufficient. A full stability analysis,
which is general and does not rely on our assumptions may be found
in \cite{ref:cr}.

We write
\begin{equation}
\omega(r,\;\theta,\;t)=\tilde{\omega}(r,\;\theta)e^{i\sigma t} , \label{eq:omegapert}
\end{equation}
and similarly for $q_{2}$ and $q_{3}$.

For time-harmonic perturbations, it can be shown that the system
governing the perturbations can be reduced to a single
second-order differential equation, which can be solved by
separating the variables $r$ and $\theta$ \cite{ref:dymg}.  In
brief, the perturbed field equations give a relation between
$\omega$ and $q_{2}$ and $q_{3}$, allowing the elimination of
$\omega$. The quantity $Q:=e^{\nu}r^{2}\sin^{3}\theta\left(
\partial_{\theta}q_{2} -\partial_{r}q_{3}\right)$ is written as
$Q=R(r)\Theta(\theta)$ and the resulting radial equation is,
asymptotically,
\begin{equation}
r^{2}e^{\nu}\frac{d}{dr} \left[\frac{e^{\nu}}{r^{2}}
\frac{dR}{dr}\right] -\mu^{2}_{l} \frac{e^{\nu}R}{r^{2}}
+\sigma^{2} R =0, \label{eq:radeq}
\end{equation}
where $\mu_{l}^{2}$ is the eigenvalue of the angular equation
which can take on the values $\mu_{l}^{2}=(l+2)(l-1)$ for
$l=2,\;3,\;...\;\;$.

We can make a change of coordinates,
\begin{equation}
r_{*}=\int e^{-\nu}\,dr\; , \nonumber
\end{equation}
so that (\ref{eq:radeq})
reduces to a Schr\"{o}dinger type equation:
\begin{equation}
\left[\frac{d^{2}}{dr^{2}_{*}} - V_{l}(r)\right]Z_{l}= -\sigma^{2}_{l}Z_{l}\label{eq:schro}
\end{equation}
where
\begin{equation}
Z_{l}(r)=\frac{R_{l}(r_{*})}{r}\;, \nonumber
\end{equation}
and
\begin{equation}
V_{l}(r)=\left(\frac{e^{\nu}}{r^{2}} \right)\left(\mu_{l}^{2}+2e^{\nu}-e^{\nu}r \nu^{\prime}\right).
\label{eq:V}
\end{equation}
The full potential, valid throughout the domain, may be found in
\cite{ref:cr} where a full stability analysis is performed. An equation of the form (\ref{eq:schro}) results
even away from the end points \cite{ref:ferrari}.

The allowed values of $\sigma$ are therefore the eigenvalues of
(\ref{eq:schro}). If all the eigenvalues are positive, then
$\sigma$ will always be real. Hence the time dependence of the
perturbations will oscillate but not grow. These perturbations
are then stable.

The eigenvalues, $\sigma_{l}$, are determined by the potential
$V_{l}(r)$. In fact, knowing the asymptotic behaviour of the
potential is sufficient for time-harmonic perturbations as
observed in \cite{ref:dymg}. The eigenvalues will be positive if
$V_{l}(r)$ is positive and the asymptotic behaviour near zero and
as $r\rightarrow \infty$ both go as $r^{-2}$.

The positivity of the full $V_{l}(r)$, as given in \cite{ref:cr},
is guaranteed if there is no DEC violation throughout the star or,
at the very least, if the DEC violation is not too large.

Regarding the asymptotics, we consider the limit as $r\rightarrow
0$ which, for gravastars with de Sitter centers gives
\begin{equation}
\lim_{r\rightarrow 0} \nu(r) = 0 \;\;\;\mbox{and}\;\;\; \lim_{r\rightarrow 0} \nu^{\prime}(r) = 0.\nonumber
\end{equation}
These conditions lead to
\begin{equation}
V_{l} \rightarrow \frac{l(l+1)}{r^2}
\end{equation}
as $r\rightarrow 0$. In the far region, $r\rightarrow \infty$, we
need to show that $\Gamma(r)$ possesses finite limit. For large
enough $r$ it can be seen (from Figure \ref{fig:8}) that $r-2m(r)
>> 1$ and that $p_{r}(r) < \rho(r)$ hence
\begin{equation}
h(r) \leq 16\pi r^{2} \rho(r). \label{eq:hrestrict}
\end{equation}
Since the integral of (\ref{eq:hrestrict}) has finite limit, $\Gamma(r)$ will also have finite limit
and therefore $\nu(r)$ also possesses finite limit. In addition. $r h(r)\rightarrow 0$ in the far zone
so that $r \nu^{\prime}(r)$ vanishes. Thus, for very large $r$, $V_{l}(r)$ has $r^{-2}$ behaviour. This
completes the argument for axial stability.

\section{{\normalsize CONCLUDING REMARKS}}
In this paper we have constructed several models of gravastars
which do not employ thin shells. These models obey the conditions
posed in \cite{[CFVis]}. The ``stellar construction'' is based on
two methods. One method employs the prescription of the energy density and
anisotropy parameter, where the system can be shown to fit
a polytropic equation of state in the stellar atmosphere. In the second method, a presciption of the energy density is imposed, and then the radial pressure is defined via a reasonable equation of state. A general model of the latter type has been shown to be stable under axial peturbations. It is expected that similar
models, utilizing the former method, would also exhibit such
stability. In closing we should mention that the models presented
here are proof-of-concept models. That is, they are constructed
to show that gravastars, under the requirements set out by
previous authors (for example \cite{[CFVis]}) can be explicitly
achieved within the framework of general relativity theory. It
would be of interest to study whether the gravitational collapse
of a heavenly body, with reasonable initial conditions and
undergoing the phase transition proposed in
\cite{[MM-1]}-\cite{ref:chls} at late times, will always form an
object without an event horizon.

\section*{{\normalsize ACKNOWLEDGEMENTS}}
D.~H. is thankful to Simon Fraser University for their hospitality and support under which a part of this work was carried out. This work was partially supported by the Ministry of Science of The Republic of Croatia under the project number ZP0036-038 (D.~H.) and the Natural Sciences and Engineering Research Council of Canada (K.~S.~V.). We would also like to thank the anonymous referees for their comments.

\linespread{0.6}
\bibliographystyle{unsrt}

\end{document}